\documentclass[preprint,12pt]{elsarticle}

\usepackage{amssymb}
\usepackage{amsmath}
\usepackage{subcaption}
\usepackage{etoolbox}
\makeatletter \emailauthor{antonio.ambrosone@unina.it}{A. Ambrosone} \emailauthor{walid.idrissiibnsalih@unicampania.it}{W. Idrissi Ibnsalih} 
\emailauthor{antonio.marinelli@unina.it}{A. Marinelli} 
\patchcmd{\emailauthor}{(#2)}{}{}{}
\emailauthor{km3net-pc@km3net.de}{}

\usepackage{lineno}
\usepackage{xcolor}
\usepackage{hyperref}
\journal{Astroparticle Physics}

\begin{document}

\begin{frontmatter}



\title{Differential Sensitivity of the KM3NeT/ARCA detector to a diffuse neutrino flux and to point-like source emission: exploring the case of the  Starburst Galaxies}





\begin{abstract}

KM3NeT/ARCA is a Cherenkov neutrino telescope under construction in the Mediterranean sea, optimised for the detection of astrophysical neutrinos with energies above $\sim$1~TeV.
In this work, using Monte Carlo simulations including all-flavour neutrinos, the integrated and differential sensitivities for KM3NeT/ARCA are presented considering the case of a diffuse neutrino flux as well as extended and point-like neutrino sources. This analysis is applied to Starburst Galaxies demonstrating that the detector has the capability of tracing TeV neutrinos from these sources. Remarkably,  after eight years, a hard power-law spectrum from the nearby Small Magellanic Cloud can be constrained. The sensitivity and discovery potential for NGC 1068 is also evaluated showing that KM3NeT/ARCA will discriminate between different astrophysical components of the measured neutrino flux after 3 years of data taking.
\end{abstract}


\cortext[cor]{corresponding author}

\author[a]{S.~Aiello}
\author[b,az]{A.~Albert}
\author[d,c]{M.~Alshamsi}
\author[e]{S. Alves Garre}
\author[c]{Z.~Aly}
\author[g,f]{A. Ambrosone\corref{cor}}
\author[h]{F.~Ameli}
\author[i]{M.~Andre}
\author[j]{E.~Androutsou}
\author[k]{M.~Anguita}
% ...

\address[a]{INFN, Sezione di Catania, (INFN-CT) Via Santa Sofia 64, Catania, 95123 Italy}
\address[b]{Universit{\'e}~de~Strasbourg,~CNRS,~IPHC~UMR~7178,~F-67000~Strasbourg,~France}
\address[c]{Aix~Marseille~Univ,~CNRS/IN2P3,~CPPM,~Marseille,~France}
\address[d]{University of Sharjah, Sharjah Academy for Astronomy, Space Sciences, and Technology, University Campus - POB 27272, Sharjah, - United Arab Emirates}
\address[e]{IFIC - Instituto de F{\'\i}sica Corpuscular (CSIC - Universitat de Val{\`e}ncia), c/Catedr{\'a}tico Jos{\'e} Beltr{\'a}n, 2, 46980 Paterna, Valencia, Spain}
\address[f]{INFN, Sezione di Napoli, Complesso Universitario di Monte S. Angelo, Via Cintia ed. G, Napoli, 80126 Italy}
\address[g]{Universit{\`a} di Napoli ``Federico II'', Dip. Scienze Fisiche ``E. Pancini'', Complesso Universitario di Monte S. Angelo, Via Cintia ed. G, Napoli, 80126 Italy}
\address[h]{INFN, Sezione di Roma, Piazzale Aldo Moro 2, Roma, 00185 Italy}
\address[i]{Universitat Polit{\`e}cnica de Catalunya, Laboratori d'Aplicacions Bioac{\'u}stiques, Centre Tecnol{\`o}gic de Vilanova i la Geltr{\'u}, Avda. Rambla Exposici{\'o}, s/n, Vilanova i la Geltr{\'u}, 08800 Spain}
\address[j]{NCSR Demokritos, Institute of Nuclear and Particle Physics, Ag. Paraskevi Attikis, Athens, 15310 Greece}
\address[k]{University of Granada, Dept.~of Computer Architecture and Technology/CITIC, 18071 Granada, Spain}
% ...
\address[az]{Universit{\'e} de Haute Alsace, rue des Fr{\`e}res Lumi{\`e}re, 68093 Mulhouse Cedex, France}
% ...

\end{frontmatter}
\section{Introduction}

The KM3NeT research infrastructure \cite{KM3Net:2016zxf} is comprised of two deep-sea Cherenkov neutrino telescopes under construction in the Mediterranean Sea: ARCA~(Astroparticle Research with Cosmics in the Abyss) located 3500 m below the sea level, off shore the coast of Capo Passero (Sicily, Italy) and ORCA (Oscillation Research with Cosmics in the Abyss)  installed at a sea-bottom depth of $\sim$2500~meters off shore Toulon (France). ARCA will have an instrumented volume of about $1~\rm Km^{3}$ and its geometry is optimised to detect high-energy astrophysical neutrinos above $\sim 1~\rm TeV$, while ORCA is designed to study  GeV neutrino physics analysing oscillation patterns of atmospheric neutrinos.
The detection principle for neutrinos\footnote{In this paper, the word neutrino is used to refer to both neutrinos and antineutrinos} in KM3NeT is based on the observation of the Cherenkov light induced in the sea water by  secondary charged particles produced in the interaction of  neutrinos inside or in the surroundings of the detector.  Given the excellent optical seawater properties, the experiment will reach unprecedented angular resolution in the identification of neutrino sources. 




In this paper, the detection performance of the full ARCA telescope, both for a  diffuse neutrino flux  and individual  sources, are explored. Using Monte Carlo simulations for all neutrino flavours, the energy-dependent 90\% confidence level (CL) sensitivity for ARCA is calculated, exploiting both track-like and shower-like events.  Track-like (referred to as \textit{tracks}) events are mainly due to charged current (CC) muon neutrino interactions;  shower-like (referred to as \textit{showers}) events are mostly due to neutral current and electron neutrinos charged current interactions.

For the track-like events, only upgoing events are considered, using the Earth as a shield to reduce the atmospheric muon contamination. Shower-like events are selected from the whole sky, using a containment requirement in the fiducial volume of the detector in order to limit the contamination from atmospheric muons and neutrinos. 
This selected sample is used to test if ARCA is able to identify TeV neutrinos from Starburst Galaxies (SBGs), either as a contribution to the diffuse flux or as emission from individual sources. 
SBGs are galaxies experiencing intense phases of star formation activity, which leads to an increased rate of supernova explosions. This enhanced activity is expected to accelerate cosmic rays up to~$\sim~\rm PeV$ energies and copiously produce gamma-rays and neutrinos~\cite{Ambrosone:2020evo,Ambrosone:2021aaw,Peretti:2019vsj}. In the diffuse flux analysis, the SBG diffuse emission up to a redshift of $\sim~4$ is considered~\cite{Ambrosone:2020evo}. In the individual source analysis, the Small Magellanic Cloud (SMC), the Circinus Galaxy \cite{Ambrosone:2021aaw} and NGC1068 \cite{IceCube:2022der} have been analysed.




The paper is organised as follows: in Sec.~\ref{sec:selection}, the event simulation and selection are outlined. In Sec.~\ref{sec:binned_lik}, the statistical analysis framework is presented. In particular, in Sec.~\ref{subsec:sensi_definition}, the definition of the sensitivity is discussed, while in Sec.~\ref{sec:diff_sensi}, the  differential limits are introduced and compared with the integrated limits. 


In Secs.~\ref{sec:diffuse_analysis} and \ref{sec:point_like},  the results of the analysis of the diffuse  neutrino flux and individual sources sensitivities are reported. In Sec.~\ref{sec:systematics}, the systematic uncertainties on the sensitivity evaluation are discussed, while in Sec.~\ref{Conclusions} conclusions are reported. Further details are shown in the accompanying appendices. In particular, \ref{sec:app_A} discusses the dependence of the sensitivity on the source extension and finally in \ref{sec:app_B}, the variation of the sensitivity in several declination bands is described.




\section{Event Simulation and Selection}\label{sec:selection}




Neutrino event interactions in the proximity of the ARCA detector 
 are generated through the gSeaGen  package~\cite{2020CoPhC.25607477A}. The atmospheric muon flux is simulated using the MUPAGE code~\cite{Carminati:2008qb}. Monte Carlo simulations continue propagating charged particles emerging from  neutrino interactions and atmospheric muon tracks through the active volume of the detector and reproducing the response of the front end electronics. The simulated data stream is filtered and reconstructed with the same algorithms used for real data. For this analysis, both upgoing tracks and all-sky showers samples are used, following the event selection discussed in~\cite{New_km3_paper}.  For the final selection, two dedicated boosted decision trees are trained in order to reject mis-reconstructed events in each sample. The  neutrino purity of the two final samples is $99\%$ for tracks and $61\%$ for showers. Moreover, the selection preserves $95\%$ of the signal neutrinos  for tracks and $70\%$ of the signal neutrinos for showers~\cite{New_km3_paper}.
The final background distribution accounts for atmospheric neutrinos and muons.

\section{Analysis Framework}\label{sec:binned_lik}

\subsection{Sensitivity Definition} \label{subsec:sensi_definition}

In this paper,~a binned maximum likelihood ratio method is used to evaluate the ARCA sensitivity. Following the formalism and the notation of~\cite{IceCube:2018fhm}, the likelihood function is defined as 
\begin{eqnarray}\label{eq:likelihood_km3}
\mathcal{L}  = \prod_i P(n_i, \mu_i)\\
\nonumber \textrm{with}\,  \mu_i = \lambda \cdot \mu_s^i + \mu_b^i
\end{eqnarray}
where $P(n,\mu)$ is the Poisson probability distribution (PDF) of observing $n$ events with  mean value $\mu$, $\mu_s$ is the expected number of signal events, while $\mu_b$  is the expected number of background events. The index $i$ runs over bins of reconstructed variables. For the following, the events are binned in reconstructed energy for the diffuse flux analysis and reconstructed energy and angular distance of the events to the source for the point-like analysis~(see below for further details). 
The signal strength $\lambda$ is left as a free normalisation parameter. Data samples are simulated by means of pseudo-experiments (PEs) generation. The procedure consists of randomly generating events according to PDFs of the signal and the background and then evaluating the test statistic $(\rm TS)$ for each PE. The $\rm TS$ is defined as
\begin{equation}\label{eq:TS_to_be_corrected}
    \rm TS = \frac{\mathcal{L}(\Tilde{\lambda})}{\mathcal{L}(\lambda = 0)}
\end{equation}
where $\Tilde{\lambda}$ is the signal strength value which maximises the likelihood for a given PE. The $90\%$ CL sensitivity is defined as the signal strength ($\lambda_{90}$) for which $90\%$ of the signal is above the median of the background-only distribution. Namely, 

\begin{equation}
    \int_{\rm TS_m} ^{+\infty} d(\rm{TS}|\lambda_{90})~d \rm{TS} = 90\%
\end{equation}
where $\rm TS_{m}$ is the median distribution in the null hypothesis (only background) and $d(\rm{TS}|\lambda_{90})$ is the PDF distribution of the TS for a given $\lambda_{90}$ value. Generally, $\lambda_{90}$ is referred to as  Model Rejection Factor (MRF), because, when signal simulations are performed according to a theoretical model, a $\rm MRF \le 1$ indicates an effective $90\%$ CL constraint on that model. On the other hand, the model discovery potential (MDP), also called the discovery flux, is defined as the minimum flux needed for a discovery with $5\sigma$ significance level in $50\%$ of the cases. 
\begin{equation}
     \int_{\rm TS_{5\sigma}} ^{+\infty} d(\rm{TS}|\lambda_{5\sigma})~d \rm{TS} = 50\%
\end{equation}
where $\rm TS_{5\sigma} = 2.85\cdot 10^{-7} $  is the $\rm TS$ threshold corresponding to a $5\sigma$ significance. This value is calculated using the one-sided Gaussian approximation \cite{KM3NeT:2018wnd}. The final sensitivity is defined as 
\begin{equation}\label{eq:final_sensi}
    \phi_{90}  (E) =  \lambda_{90}  \phi_s(E)
\end{equation}
while the discovery flux as
\begin{equation}\label{eq:final_MDP}
    \phi_{5\sigma}  (E) =  \lambda_{5\sigma}  \phi_s(E)
\end{equation}
where $\phi_s (E)$ is the signal flux injected in the simulations corresponding to $\lambda =1$. Typically, the injected signal spectrum $\phi_s(E)$ is modelled as a power-law $E^{-\gamma}$, resulting in sensitivity and MDP profiles mirroring the shape of the injected spectrum. In fact, Eqs. \ref{eq:final_sensi} and \ref{eq:final_MDP} provide results depending solely on the spectral shape and are entirely independent of the specific scaling or normalisation of the injected signal. Such expectations are also referred to as energy-integrated expectations, see~\cite{New_km3_paper}.



\subsection{Differential Sensitivity and Discovery Flux}\label{sec:diff_sensi}

The differential sensitivity and discovery flux  (in general, referred to as `differential limits') correspond to the performance of a detector for a given energy range~\cite{IceCube:2018fhm}.
In other words, they represent the minimum differential signal neutrino flux that the detector can either constrain or discover. 
They are only mildly model-dependent and represent important instrument response functions. In order to evaluate them, the signal is divided in bins of logarithm of true energy, injecting an $E^{-2}$ spectrum within each bin. The width of the bins is chosen to be half decade in energy. Since the sensitivity is inversely proportional to the number of signal events $n_{s}$ provided by the injected spectrum~\cite{Feldman:1997qc}, there is a strict connection between differential and integrated sensitivities. In particular, 
\begin{equation}\label{eq:int_vs_diff}
    \lambda_{90} \propto \rm n_s^{-1} = \frac{1}{\sum_{i}n_s^{i}} \propto \frac{1}{\sum_{i}(\lambda_{90}^{i})^{-1}}
\end{equation}
where $\lambda_{90}$ is the integrated MRF, \textit{i} runs over the number of energy bins and $\lambda_{90}^{i}$ is the MRF for each energy bin of the signal. \newline 
The integrated MRF, as shown in Eq.~\ref{eq:int_vs_diff},  is always smaller than the differential MRF. In fact, in differential limits, each individual energy bin is treated as independent, assuming a zero signal flux outside the bin. This results in the same amount of  background events within each bin, leading to worse limits. The integrated sensitivity exploits the energy dependence of the flux to distinguish between the signal and the background, while the differential sensitivity provides the generic differential flux which the detector can discriminate, leading to a more stringent requirement. On the other hand, the differential limits better highlight the energy range where the detector is most sensitive, especially for different event samples (tracks and showers). 
\newline
In the following, these quantities are computed and discussed in the context of the sensitivity of ARCA to neutrino emission from starburst galaxies.



\section{Diffuse Analysis}\label{sec:diffuse_analysis}

\begin{figure}[h!]
    \centering
    \includegraphics[width=\columnwidth]{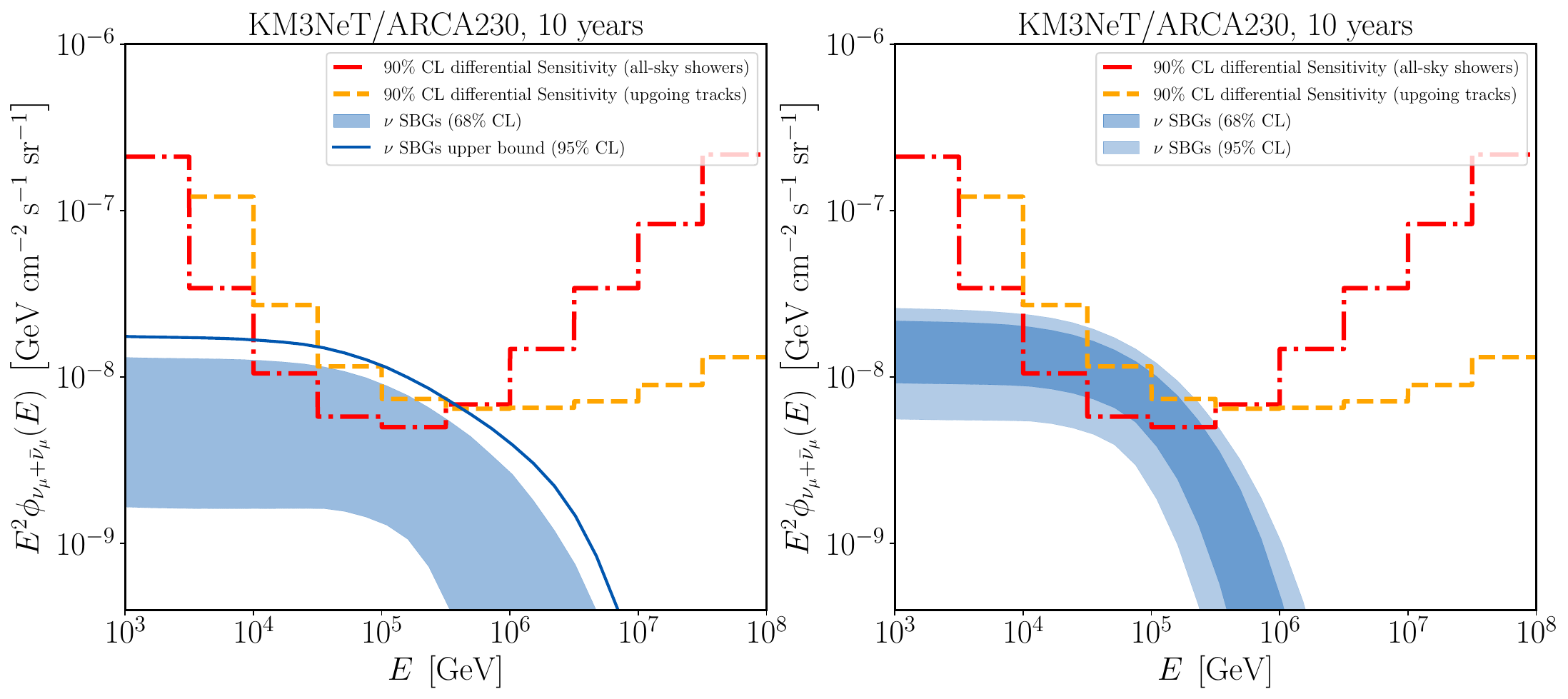}
    \caption{$90\%$ CL differential sensitivity to a diffuse neutrino flux for the sample of upgoing tracks (dashed orange) and of all-sky showers (dashed-dotted red). On the left, the sensitivities are compared to the theoretical $1\sigma$ band prediction for SBGs  neutrino as obtained in~\cite{Ambrosone:2020evo} through a multi-component fit of the extragalactic gamma-ray background (EGB) measured by Fermi-LAT \cite{Fermi-LAT:2014ryh}, and the 7.5~yr HESE neutrino flux measured by IceCube \cite{IceCube:2020wum}. On the right, the KM3NeT/ARCA expectations are compared to the blue band which corresponds to the  $1\sigma$  SBG  neutrino expectations from  \cite{Ambrosone:2020evo} obtained by a multi-component fit of the EGB measured by Fermi-LAT \cite{Fermi-LAT:2014ryh} and the 6-yr shower neutrino flux measured by IceCube \cite{IceCube:2020acn}.}
    \label{fig:diffuse_plot}
\end{figure}

The $90\%$ CL differential sensitivity for a diffuse neutrino flux both for the upgoing tracks (orange dashed line) and all-sky showers (red dashed-dotted line) is shown in Fig. \ref{fig:diffuse_plot}.
For comparison, the $1~\sigma$ bands around the diffuse neutrino flux from SBGs predicted by~\cite{Ambrosone:2020evo} from SBGs are also reported.
Predictions shown in the left (right) panel are obtained through a multi-component fit  of the extragalactic gamma-ray background measured by Fermi-LAT~\cite{Fermi-LAT:2014ryh} and the 7.5 yr HESE neutrino flux~\cite{IceCube:2020wum} (6 yr shower neutrino flux~\cite{IceCube:2020acn}) measured by IceCube (see~\cite{Ambrosone:2020evo} for further details).

\noindent
The sensitivities refer to 10 years of data collected by the ARCA detector for one single flavour of neutrinos.
For neutrino energies below $100\, \, \rm TeV$, the sensitivity is predominantly driven by purely upgoing events $(\theta \le 80^{\circ})$, where $\theta$ is the reconstructed zenith angle of the tracks and $\theta =0^{\circ}$ corresponds to a vertical upgoing track. For energies exceeding $100\, \,  \rm TeV$, the sensitivity is dominated by  horizontal events $(80^{\circ}<\theta <100^{\circ})$ due to neutrino absorption in the Earth. This  results in a minimum of the sensitivity around $1\, \rm PeV$.
The minimum of the sensitivity for showers is at~$\simeq 100\,\rm TeV$ because the event containment significantly reduces the amount of observed signal above this energy.  Due to a reduced background rate,  the shower sensitivity is better than the one for tracks for $\rm E_{\nu} \lesssim 100\, \rm TeV$. For $\rm E_{\nu}\gtrsim 100\, \rm TeV$, the sensitivity is dominated by the tracks,  because of a larger effective volume for track-like events as the energy increases. After 10 years of operation, ARCA will probe the neutrino emission from SBGs, either confirming or constraining their contribution to the diffuse neutrino flux around~$\sim 100\,\rm TeV$. 
The sensitivities shown in  Fig.~\ref{fig:diffuse_plot} suggest that ARCA will also improve the characterisation of the diffuse neutrino flux provided by IceCube~\cite{IceCube:2020wum,IceCube:2020acn}, thus providing valuable information regarding other astrophysical sources contributing to the diffuse neutrino flux such as gamma-ray opaque sources~\cite{Fang:2022trf}.






\section{Point-like SBG Analysis}\label{sec:point_like}
ARCA is  expected to have an excellent angular resolution ($\lesssim 0.2^{\circ}$ for tracks and $\rm E_{\nu} \ge 10\, \rm TeV$ see~\cite{New_km3_paper}, and therefore an enhanced sensitivity for point-like and extended sources. The likelihood function takes into account both the reconstructed energy $(\rm E_{\rm rec})$ and the angular distance between the nominal position of the source and the reconstructed direction of the event within an angle $\alpha$. The range $[10,10^{8}]\,~\rm GeV$ and $[0,5^{\circ}]$ $([0,15^{\circ}])$ for tracks (showers) are respectively considered for the reconstructed energy and $\alpha$.
In the following, the evaluation of the differential limits is performed considering the position of three local SBGs in order to test the ability of ARCA to trace their star-forming activity through neutrino emission. Specifically, the Small Magellanic Cloud  and the Circinus Galaxy are studied, because their expected neutrino fluxes should be high enough to match ARCA sensitivity after $\sim 6$ years of operation \cite{Ambrosone:2021aaw}. The case of NGC~1068, whose neutrino emission has been measured by IceCube at $4.2\sigma$ significance with a neutrino flux normalisation $\phi_{\nu_{\mu}+\bar{\nu}_{\mu}} = 5.0 \times 10^{-11}\, \rm TeV^{-1}\, \rm cm^{-2}\, \rm s^{-1}$ at $\rm E_{\nu} = 1\, \rm TeV$~\cite{IceCube:2022der}, is also considered.





\subsection{Small Magellanic Cloud}

The Small Magellanic Cloud  is a local star-forming galaxy at a distance of $\sim$ $60\, \rm{kpc}$ \cite{Kornecki:2020riv}, characterised by a stable  and diffuse gamma-ray emission, driven by its star-forming activity~\cite{Ajello:2020zna}. Its expected neutrino emission has been estimated using a one-zone model tuned to its gamma-ray spectrum by \cite{Ambrosone:2021aaw}. The capability of ARCA to test this model is evaluated by simulating  the SMC as  an extended source with a radius $r = 0.5^{\circ}$, consistent with its angular extension measured by the Fermi-LAT Collaboration \cite{Ajello:2020zna}.
\begin{figure}[h!]
    \centering
    \includegraphics[width=0.49\columnwidth]{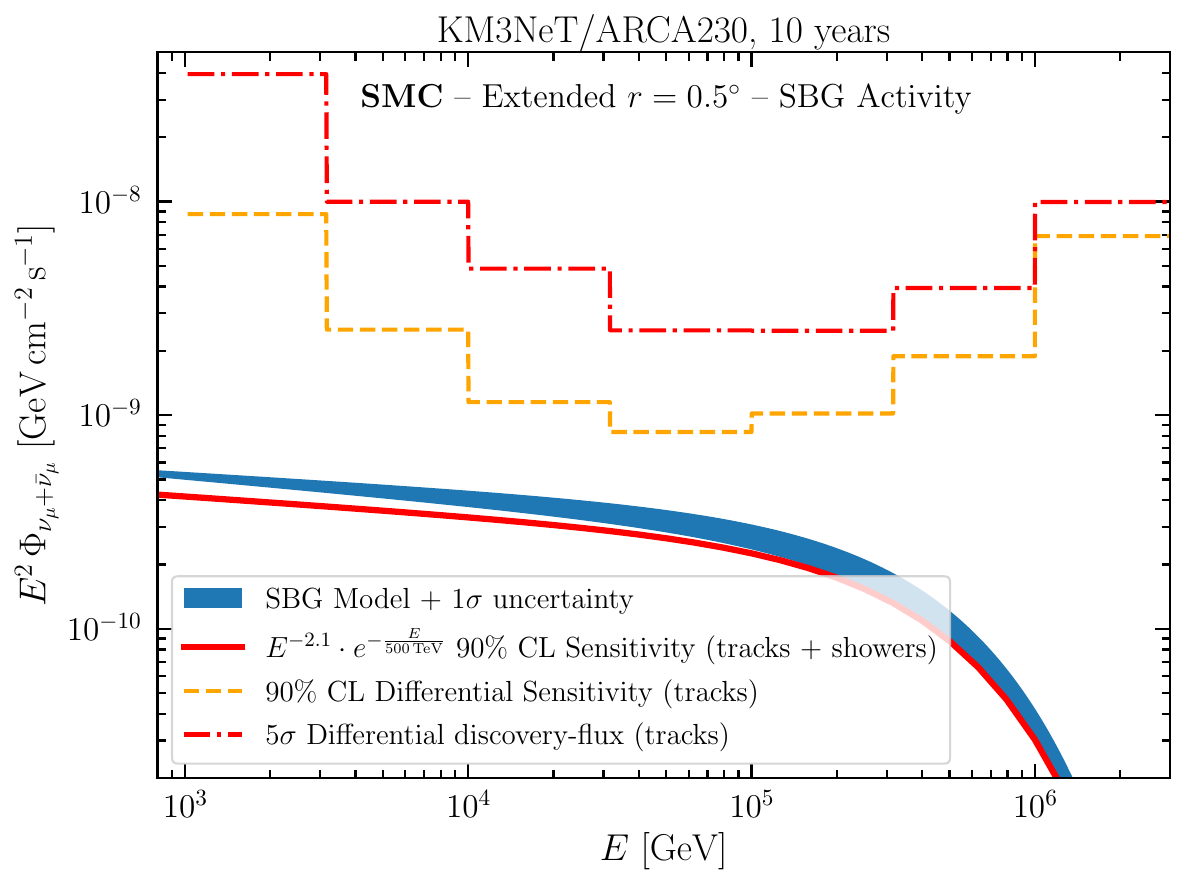}
    \includegraphics[width=0.49\columnwidth]{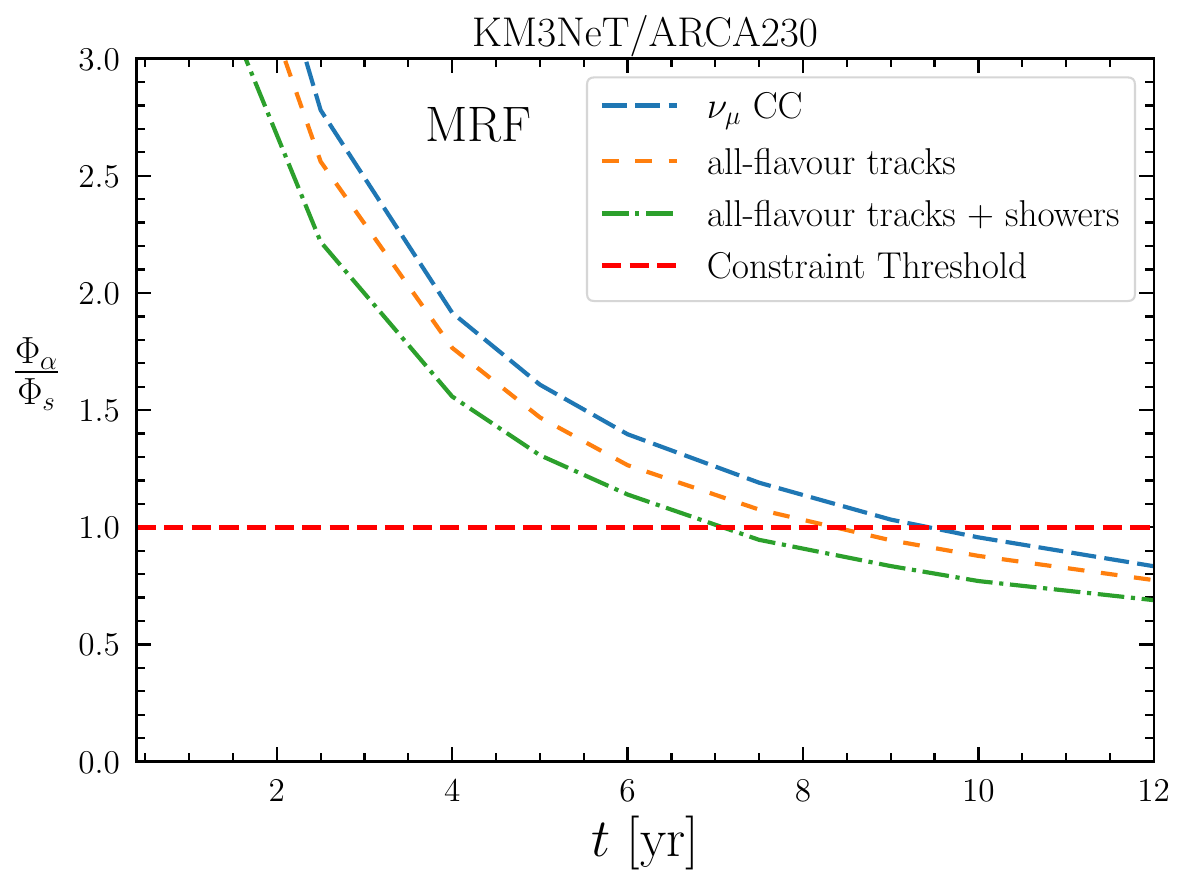}
    \caption{Left: $90\%$  differential sensitivity  (orange dashed line) and the $5\sigma$  differential discovery flux (red dashed-dotted line) of ARCA after 10 years of operation for one neutrino flavour compared with the $1\sigma$ SBG neutrino model (blue band) from \cite{Ambrosone:2021aaw}. The energy-integrated sensitivity (tracks+showers) for the best-fit spectrum of the source (considering 10 years of operation) is also reported. Right: energy-integrated MRF (best fit flux) as a function of time for different event samples. The blue line refers to $\nu_{\mu}$ CC events, orange to the all-flavour tracks, and green to tracks+showers. }
    \label{fig:SMC_results}
\end{figure}

The $90\%$ CL  differential sensitivity and the $5\sigma$  differential discovery flux for the SMC, considering 10 years of ARCA operation, are shown in the left panel of Fig.~\ref{fig:SMC_results} and compared with the $1\sigma$ neutrino expectations  from  \cite{Ambrosone:2021aaw}. The track+shower energy-integrated sensitivity is also reported  for the best-fit spectrum of the source which can be described by a power-law with an exponential cutoff $\sim \rm E^{-2.1}\cdot \rm e^{(-\rm E/500\, \rm TeV)}$. The cutoff at $500\,\rm TeV$ comes from the assumption of maximal proton energy~$\simeq 10\,\rm PeV$. In the right panel of Fig.~\ref{fig:SMC_results}, The energy-integrated MRF is shown as a function of time. Three different event samples are considered: $\nu_{\mu}$ in  CC, all-flavour tracks~($\nu_{\mu}$ CC interactions and $\nu_{\tau}$  in CC interactions where the $\tau$ decays into a muon) and finally all-flavour tracks +showers. The horizontal threshold line $(\rm MRF = 1)$ is superimposed for comparison. 
Although the detector is not sensitive to discriminate the flux in single true energy signal bins, the integrated spectrum can be constrained after~$\sim 8\, \rm years$, combining the information from the track-like and shower-like sample. The samples provide different results due to a diverse amount of events (signal and background) contained in each sample.
Therefore, this will test the model proposed by \cite{Ambrosone:2021aaw}. It is important to stress that this source, having a low star formation rate, might have a sizeable contamination from leptonic point-like sources, such as pulsar wind nebulae, in its gamma-ray spectrum, leading to  a smaller neutrino spectrum than the one provided in  \cite{Ambrosone:2021aaw}. 
This will allow for a crucial test of the neutrino content of the source.
If the neutrino emission is limited in a smaller region than the entire galaxy, the sensitivity improves, as discussed in~\ref{sec:app_A}.





\subsection{NGC 1068}

NGC 1068 is a source situated at $\sim$10-14~Mpc from the Milky Way \cite{Ajello:2020zna}. It is classified as a Starburst and as a type II Seyfert galaxy, characterised by a star formation rate of $\sim$ $23\, \rm M_{\odot}\, \rm yr^{-1}$ and by the presence of a starburst ring \cite{Eichmann:2022lxh}. Its Seyfert activity is located mainly inside its active nucleus \cite{Ajello:2020zna,Antonucci:1985aa}, and its hot corona ~\cite{Ajello:2020zna,Eichmann:2022lxh,Inoue:2019yfs,Kheirandish:2021wkm,Anchordoqui:2021vms,Murase:2019vdl}.  The ALMA telescope \cite{Eichmann:2022lxh, 2006AJ....132..546G,Michiyama:2022nsx}
has also detected a powerful radio jet and four parsec-scale blobs in the jet head, which may represent sites for cosmic ray  acceleration. Furthermore, a hint of an ultra-fast outflow in the core of the source has been reported by \cite{Mizumoto:2018lfe} analysing absorption lines in the X-ray band. Recently, the  IceCube Collaboration has reported an excess of {$79^{+22}_{-20}$ astrophysical neutrinos from the direction of NGC 1068,  with an overall significance of $4.2\sigma$ above the background-only hypothesis \cite{IceCube:2022der}.  The measured neutrino flux normalisation is at least a  factor 10 above the measured gamma-ray flux~\cite{Ajello:2020zna,MAGIC:2019fvw}, rather than the typical factor of about 2 expected from hadronic emission in transparent sources. 
\newline
Both AGN and SBG activities have been used to explain the origin of such a high neutrino flux. However, the SBG-related emission produces  a neutrino flux approximately two orders of magnitude below the IceCube measurements~\cite{Ambrosone:2021aaw,Merckx:2023kvn}.  On the other hand, AGN-related emissions can produce a greater contribution to observed neutrino flux~~\cite{Peretti:2023xqk,Fang:2023vdg,Marinelli:2023ov}. At the moment, the favoured scenario seems a considerable neutrino emission related to the magnetised hot corona activity~\cite{Kheirandish:2021wkm}.

The time-integrated expectations for ARCA are shown in Fig.~\ref{fig:NGC_results}.

\begin{figure}[h!]
    \centering
    \includegraphics[width=0.495\columnwidth]{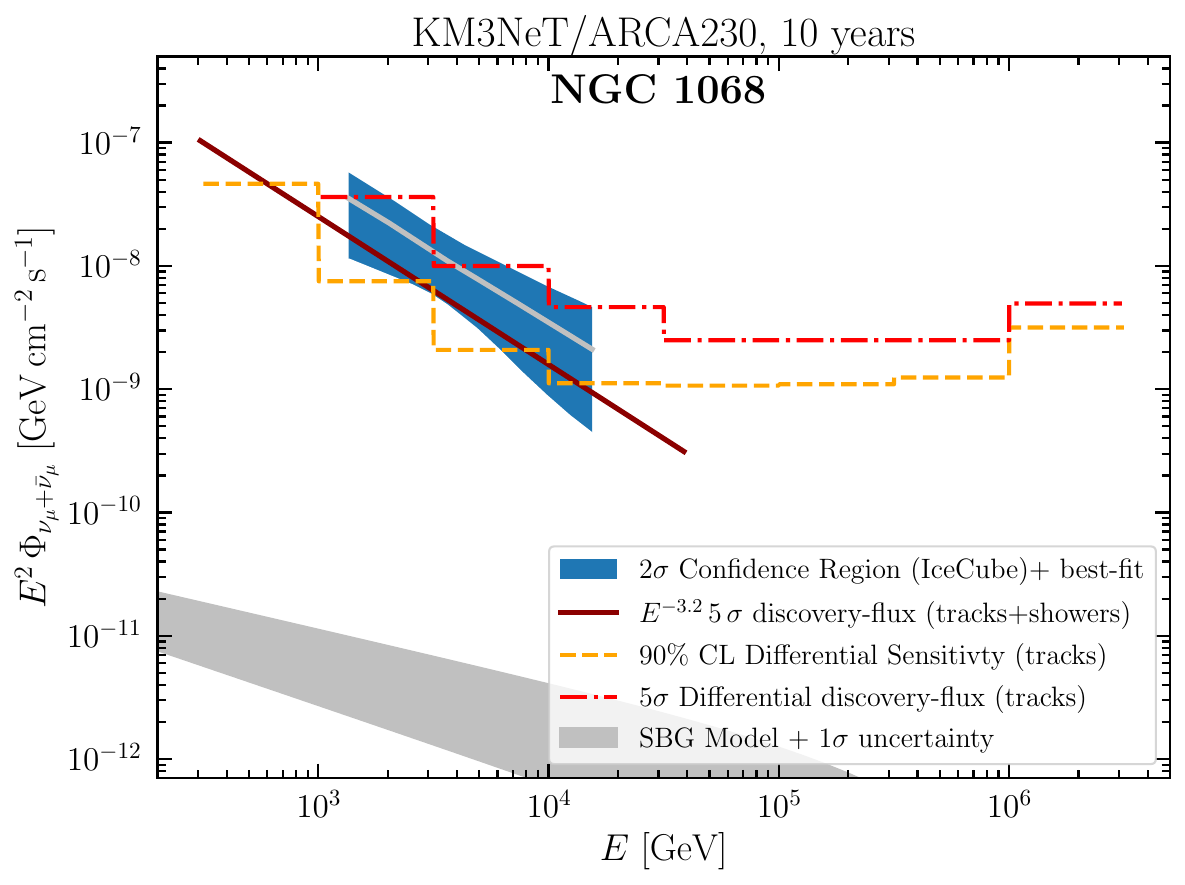}
    \includegraphics[width=0.495\columnwidth]{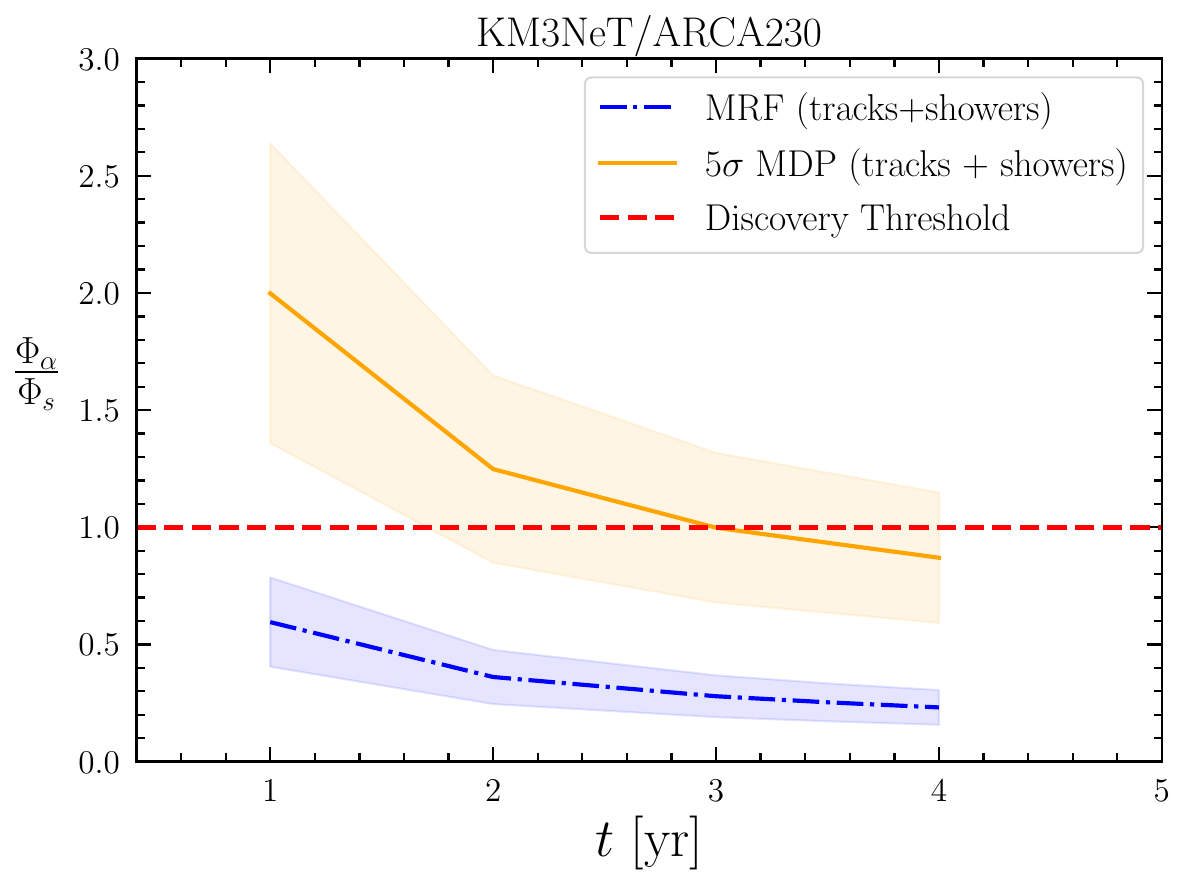}
    \caption{Left: $90\%$ CL differential sensitivity (orange dashed line) and $5\sigma$ differential discovery flux (red dashed-dotted line) and the energy-integrated $5\sigma$ discovery for an $E^{-3.2}$ spectrum using the tracks+showers sample (dark red line). The expectations are compared with the IceCube-measured $2\sigma$ region for the spectrum~\cite{IceCube:2022der} (blue band) and the theoretical neutrino predictions from SBG activity according to \cite{Ambrosone:2021aaw} (grey band). Right: Energy-integrated $5\sigma$ discovery flux and $90\%$ CL sensitivity as functions of the operation time based on the combined track+shower sample. The bands correspond to the the uncertainty of the flux measured by~\cite{IceCube:2022der}.
    After $\sim 3\, \rm years$, KM3NeT/ARCA is expected to confirm the spectrum measured by IceCube.}
    \label{fig:NGC_results}
\end{figure}

\noindent
In the left panel of Fig.~\ref{fig:NGC_results}, the $90\%$ CL differential sensitivity and the $5\sigma$  discovery flux for $10\, \rm years$ of data-taking for ARCA are compared with the IceCube $2\sigma$ region~\cite{IceCube:2022der} (blue band). The obtained differential limits show that ARCA can probe a neutrino flux of $E^2\phi_{\nu_{\mu} +\bar{\nu}_{\mu}}\sim 5\cdot 10^{-8}\, \rm GeV\, \rm cm^{-2}\, \rm s^{-1}$ and of $E^2\phi_{\nu_{\mu} +\bar{\nu}_{\mu}}\sim 10^{-9}\, \rm GeV\, \rm cm^{-2}\, \rm s^{-1}$ at $1\,\ \rm TeV$ and $100\, \rm TeV$, respectively.
The energy-integrated 5$\sigma$ discovery for a $E^{-3.2}$ spectrum  and the expected SBG neutrino spectral energy distribution evaluated according to~\cite{Ambrosone:2021aaw}  are also shown for comparison. 

In the right panel of Fig.~\ref{fig:NGC_results}, the energy-integrated discovery flux and MRF are reported as functions of observation time and compared with the discovery threshold corresponding to~$\rm MDP=1$.
The bands refers to the uncertainty of the flux measured by~\cite{IceCube:2022der}.

The ARCA detector is expected to confirm the flux reported as a best fit by IceCube after only $\sim 3\, \rm years$ of data taking and to better disentangling the emission from AGN-related components from the SBG-related one.

\subsection{Circinus Galaxy}
Circinus is a SBG located $\sim$ $4\, \rm Mpc$ from the Milky Way \cite{Kornecki:2020riv} at a declination $\delta = -65.2^{\circ}$ \cite{2023A&A...670A..18O}, where the ARCA detector is expected to have full visibility \cite{KM3NeT:2018wnd}. Circinus is classified as a Seyfert~II galaxy \cite{Maiolino_1998}, showing both AGN activity and a hot corona \cite{Kheirandish:2021wkm,Stalevski_2017,Murase:2023ccp}. The neutrino flux emission predicted by the hot corona model can reach a flux level of $E^{2}\phi_{\nu_{\mu} + \bar{\nu}_{\mu}} \simeq 10^{-8}\, \rm GeV \, \rm cm^{-2}\, \rm s^{-1}$ \cite{Kheirandish:2021wkm} above $\rm E_{\nu} \sim 1\, \rm TeV$.

\begin{figure}[h!]
    \centering
    \includegraphics[width=0.49\columnwidth]{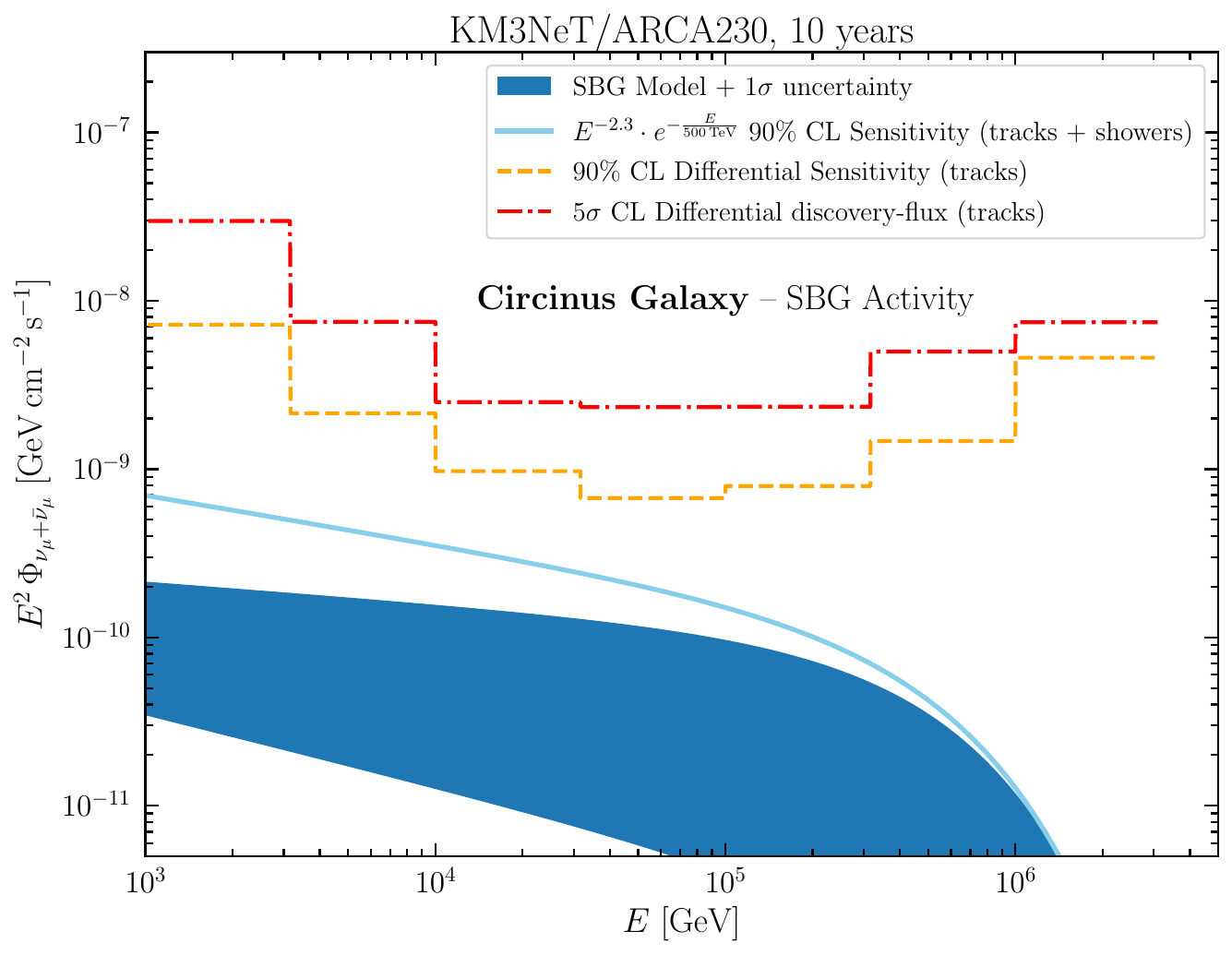}
    \includegraphics[width=0.49\columnwidth]{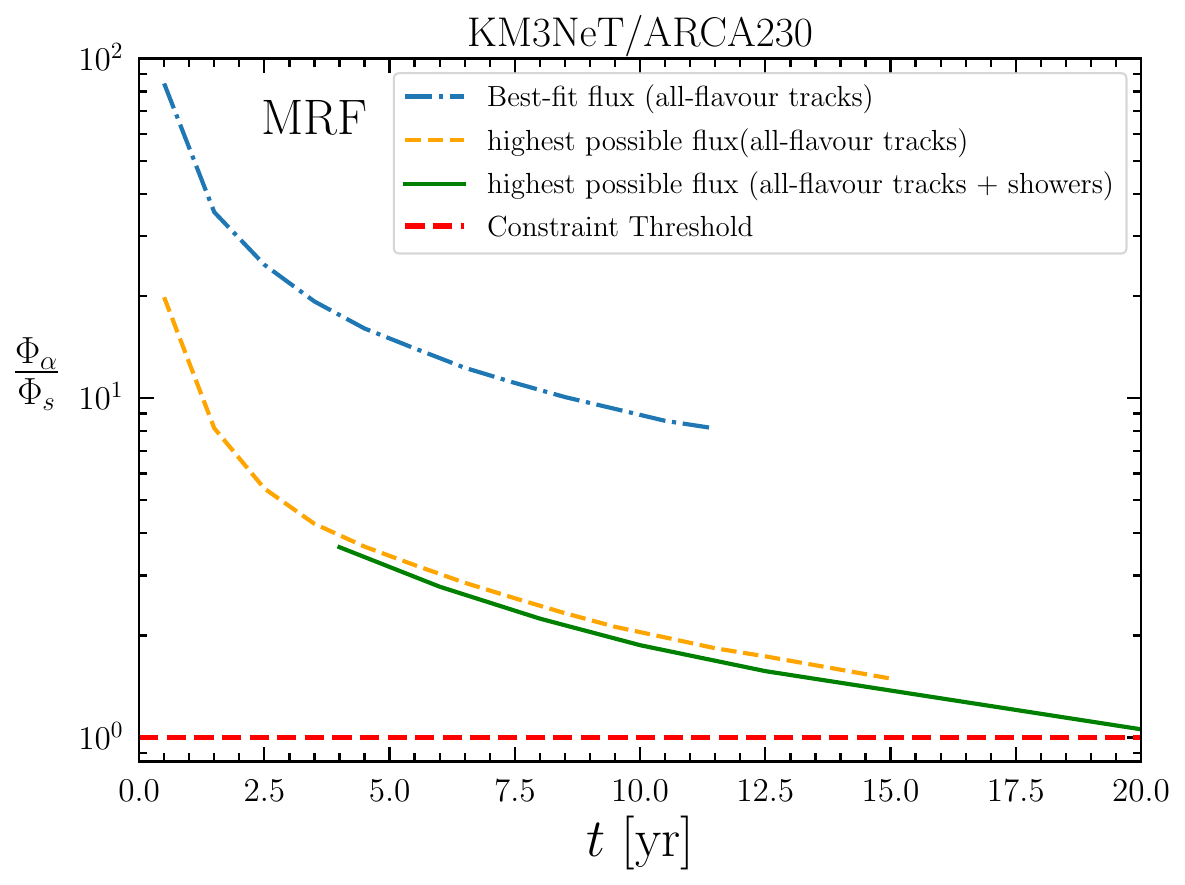}
    \caption{Left: $90\%$ CL track differential sensitivity (dashed orange line), $5\sigma$ track differential discovery flux (dashed-dotted red line) for 10 years of data taking with the ARCA detector, compared with the $1\sigma$ SBG neutrino flux predictions from~\cite{Ambrosone:2021aaw} (blue band). Also shown is the energy-integrated $90\%$ CL sensitivity for $E^{-2.3}\cdot \rm e^{-\rm E/500\, \rm TeV}$, corresponding to the best-fit neutrino spectrum from SBG activity. Right: The MRF as a function of observation time for the best-fit spectrum of the source (dashed-dotted blue line, all-flavour tracks), for the highest possible flux $(1\sigma)$ (dashed orange line, all-flavour tracks) and for the highest possible flux considering tracks and showers (green line).}
    \label{fig:Circinus_results}
\end{figure}


\noindent
In the left panel of Fig.~\ref{fig:Circinus_results}, the $90\%$ CL  differential sensitivity and the $5\sigma$  differential discovery flux are compared with the $1\sigma$ neutrino expectations from SBG activity~\cite{Ambrosone:2021aaw}.
The differential limits indicate that a flux of the order of $E^{2}\phi_{\nu_{\mu}+\bar{\nu}_{\mu}} \simeq 10^{-9}\, \rm GeV \rm cm^{-2}\, \rm s^{-1}$ can be probed in the energy range $30-100\, \rm TeV$. Therefore, a neutrino flux connected to the AGN activity can be searched for with ARCA, which may lead to the discovery of a  hot corona emission like in NGC 1068. 
The energy-integrated sensitivity for the best-fit SBG flux~$(\rm E^{-2.3}\cdot e^{(-\rm E/500\, \rm TeV)})$ is also reported. In the right panel of Fig.4, the MRF is presented as a function of observation time for different cases: the best-fit flux of the source analysed  with all-flavours tracks and the highest possible flux $(1\sigma)$ analysed with all-flavour tracks and all-flavour tracks + showers. The obtained sensitivity shows that the latter can be constrained after $\sim 20\,$ years of observations.





\section{Impact of Systematic Uncertainties}\label{sec:systematics}
In this section, a brief discussion on the systematic uncertainties associated with the ARCA sensitivity is presented.
The most significant factors impacting the sensitivity are a $25\%$ uncertainty in the conventional atmospheric neutrino flux and a $10\%$ uncertainty in the effective area~\cite{KM3Net:2016zxf}. The resulting uncertainty on the sensitivity from these effects can be conservatively estimated by using the fact that, in background-dominated data, $\rm MRF \propto \sqrt{\rm n_{back}}/\rm n_s$, where $\rm n_{back}$ is the expected number of background events and $\rm n_s$ is the expected number of signal events provided by the input spectrum~\cite{Feldman:1997qc}.
Since both the numbers of events (signal and background) are proportional to the effective area, a variation of 10\% leads to a $4.6\%$ uncertainty on the sensitivity. Furthermore, conservatively assuming that the background sample is dominated by atmospheric neutrinos, 
an uncertainty of $25\%$ on the background rate leads to a $12.5\%$ uncertainty on the sensitivity.
Regarding the uncertainty of the discovery flux, Ref.~\cite{KM3NeT:2018wnd} has shown that the $25\%$ uncertainty on the atmospheric neutrino flux leads to  a $+15\%$ and $-5\%$ uncertainty. Therefore, it is expected that for the SMC, the time required to achieve a 90\% CL constraint varies within $8.0^{+2.2}_{-1.7}, \rm years$, whereas for the highest possible flux from the Circinus galaxy, it is $20\pm 5, \rm years$. In contrast, the discovery of the $E^{-3.2}$ spectrum observed by IceCube in the direction of NGC 1068 yields an estimated time of $3.0^{+1.0}_{-0.3}, \rm yr$.

\section{Final Remarks and Conclusions}\label{Conclusions}
\noindent




In this paper, the expected differential sensitivity for the ARCA detector is presented for the first time for a diffuse neutrino flux and individual neutrino sources. A binned maximum likelihood formalism is applied to evaluate the $90\%$ CL differential sensitivity and the $5\sigma$ differential discovery flux, by binning the signal flux in true energy bins of half-decade widths. The differential limits are shown to be mostly signal model-independent and complementary to the energy-integrated limits and provide crucial information regarding the energy ranges where the detector is more sensitive to neutrino signals. They are evaluated using both upgoing track-like and all-sky contained shower-like events.
\newline
The results show that ARCA will  probe the shape of the diffuse flux for $E \lesssim 100\, \rm TeV$,  constraining the role of sources such as SBGs and complementing the IceCube's observations of the neutrino sky. 
\newline
The likelihood formalism  is also applied to local SBG sources, following the flux predictions provided in~\cite{Ambrosone:2021aaw}. After 8 years of data taking, the Small Magellanic Cloud can be observed as a $90\%$ CL excess in the energy-integrated analysis, proving that SBGs are guaranteed neutrino emitters. ARCA expectations for NGC~1068 have been evaluated indicating that ARCA will need only$\sim$3~years for discovering this source with $5\sigma$ significance. This will better disentangle the AGN-related contributions from the SBG-related one, strengthening IceCube observations.
 Finally, the differential limits for the Circinus Galaxy are evaluated, showing that the detector can probe its potential AGN-related emission activity additional to the expected SBG neutrino flux.





\section{Acknowledgements}

The authors gratefully acknowledge the contribution from Damiano F. G. Fiorillo and Marco Chianese for providing support regarding SBG fluxes and for fruitful discussions.

The authors acknowledge the financial support of the funding agencies:

Czech Science Foundation (GAČR 24-12702S);

Agence Nationale de la Recherche (contract ANR-15-CE31-0020), Centre National de la Recherche Scientifique (CNRS), Commission Europ\'eenne (FEDER fund and Marie Curie Program), LabEx UnivEarthS (ANR-10-LABX-0023 and ANR-18-IDEX-0001), Paris \^Ile-de-France Region, France;

Shota Rustaveli National Science Foundation of Georgia (SRNSFG, FR-22-13708), Georgia;

The General Secretariat of Research and Innovation (GSRI), Greece;

Istituto Nazionale di Fisica Nucleare (INFN) and Ministero dell’Universit{\`a} e della Ricerca (MUR), through PRIN 2022 program (Grant PANTHEON 2022E2J4RK, Next Generation EU) and PON R\&I program (Avviso n. 424 del 28 febbraio 2018, Progetto PACK-PIR01 00021), Italy; A. De Benedittis, R. Del Burgo, W. Idrissi Ibnsalih, A. Nayerhoda, G. Papalashvili, I. C. Rea, S. Santanastaso, A. Simonelli have been supported by the Italian Ministero dell'Universit{\`a} e della Ricerca (MUR), Progetto CIR01 00021 (Avviso n. 2595 del 24 dicembre 2019);

Ministry of Higher Education, Scientific Research and Innovation, Morocco, and the Arab Fund for Economic and Social Development, Kuwait;

Nederlandse organisatie voor Wetenschappelijk Onderzoek (NWO), the Netherlands;

The National Science Centre, Poland (2021/41/N/ST2/01177); The grant “AstroCeNT: Particle Astrophysics Science and Technology Centre”, carried out within the International Research Agendas programme of the Foundation for Polish Science financed by the European Union under the European Regional Development Fund;

National Authority for Scientific Research (ANCS), Romania;

Slovak Research and Development Agency under Contract No. APVV-22-0413; Ministry of Education, Research, Development and Youth of the Slovak Republic;

MCIN for PID2021-124591NB-C41, -C42, -C43 funded by MCIN/AEI/ 10.13039/501100011033 and, as appropriate, by “ERDF A way of making Europe”, by the “European Union” or by the “European Union NextGenerationEU/PRTR”, Programa de Planes Complementarios I+D+I (refs. ASFAE/2022/023, ASFAE/2022/014), Programa Prometeo (PROMETEO/2020/019) and GenT (refs. CIDEGENT/2018/034, /2019/043, /2020/049. /2021/23) of the Generalitat Valenciana, Junta de Andaluc\'{i}a (ref. SOMM17/6104/UGR, P18-FR-5057), EU: MSC program (ref. 101025085), Programa Mar\'{i}a Zambrano (Spanish Ministry of Universities, funded by the European Union, NextGenerationEU), Spain;
The European Union's Horizon 2020 Research and Innovation Programme (ChETEC-INFRA - Project no. 101008324).

\clearpage

\bibliography{references}
\bibliographystyle{unsrt}




\appendix

\section{Impact of Extension to the Differential Sensitivity}\label{sec:app_A}
The differential sensitivity at the declination of the Small Magellanic Cloud after 10 years of operation of the ARCA detector (for the tracks) is shown in Fig.~\ref{fig:Extension_sensitity} considering a point-like source and an extended source with $r = 0.5^{\circ}$.
\begin{figure}[h!]
    \centering
    \includegraphics[width=\columnwidth]{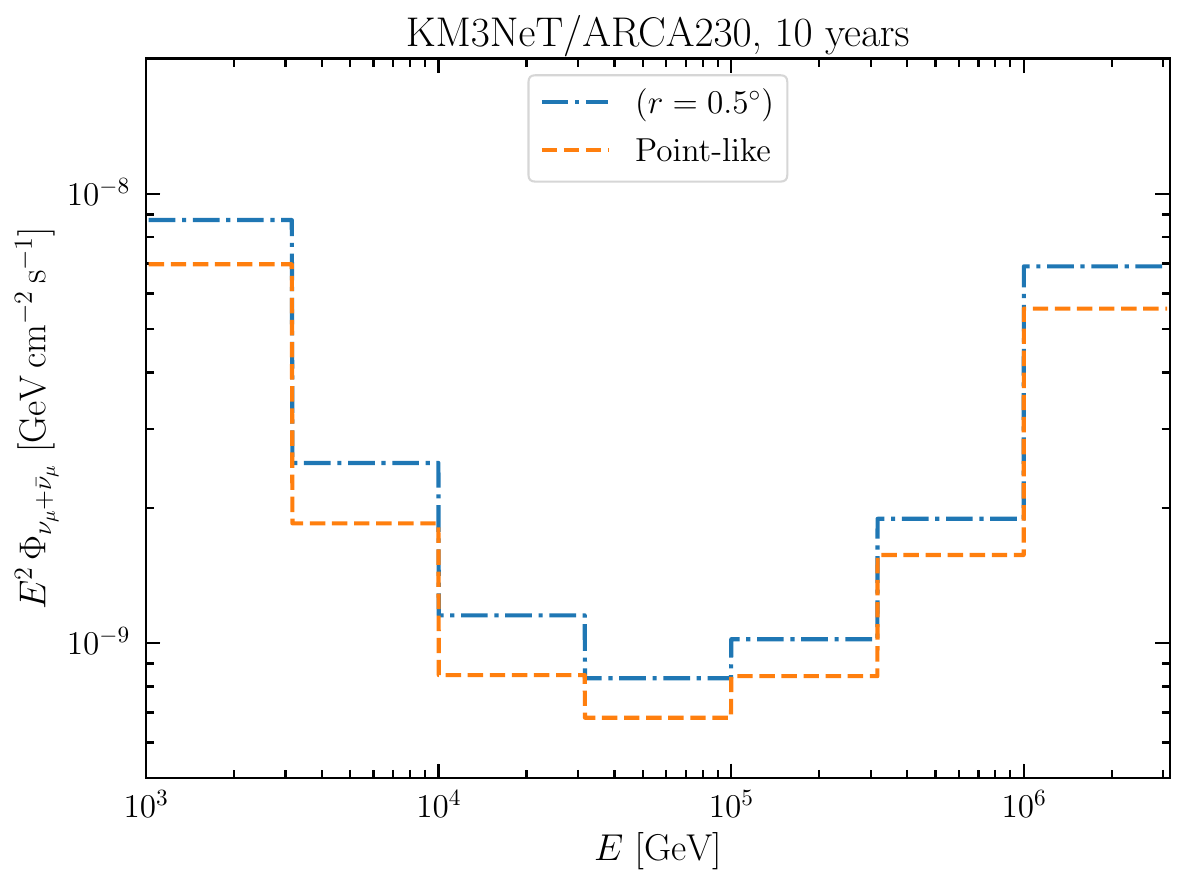}
    \caption{Comparison between the $90\%$ CL differential sensitivity for a point-like source (orange dashed line) and for an extended source with an extension of $r = 0.5^{\circ}$~(blue dashed-dotted line), considering  10 years of data-taking at the declination of the SMC.  }
    \label{fig:Extension_sensitity}
\end{figure}
The differential sensitivity of the extended source worsens by $\sim$$30\%$ with respect to the point-like source case. Therefore, in case the neutrino emission for SBGs is concentrated within their nuclei,  the sensitivity is expected to be close to the one expected for point-like sources  and the time required for a $90\%$ CL excess might reduce with the one presented in the main text.

\section{Sensitivity Dependence on Declination}\label{sec:app_B}
The comparison  between the differential sensitivities after 10 years of operation of ARCA is shown in Fig.~\ref{fig:sensi_declination} for three declinations:  $\delta = -73^{\circ}$, $\delta = -65.3^{\circ}$ and finally $\delta = -0.01^{\circ}$. These declinations correspond to the positions of the SMC, Circinus Galaxy, and NGC 1068, respectively.
\begin{figure}[h!]
    \centering
    \includegraphics[width=\columnwidth]{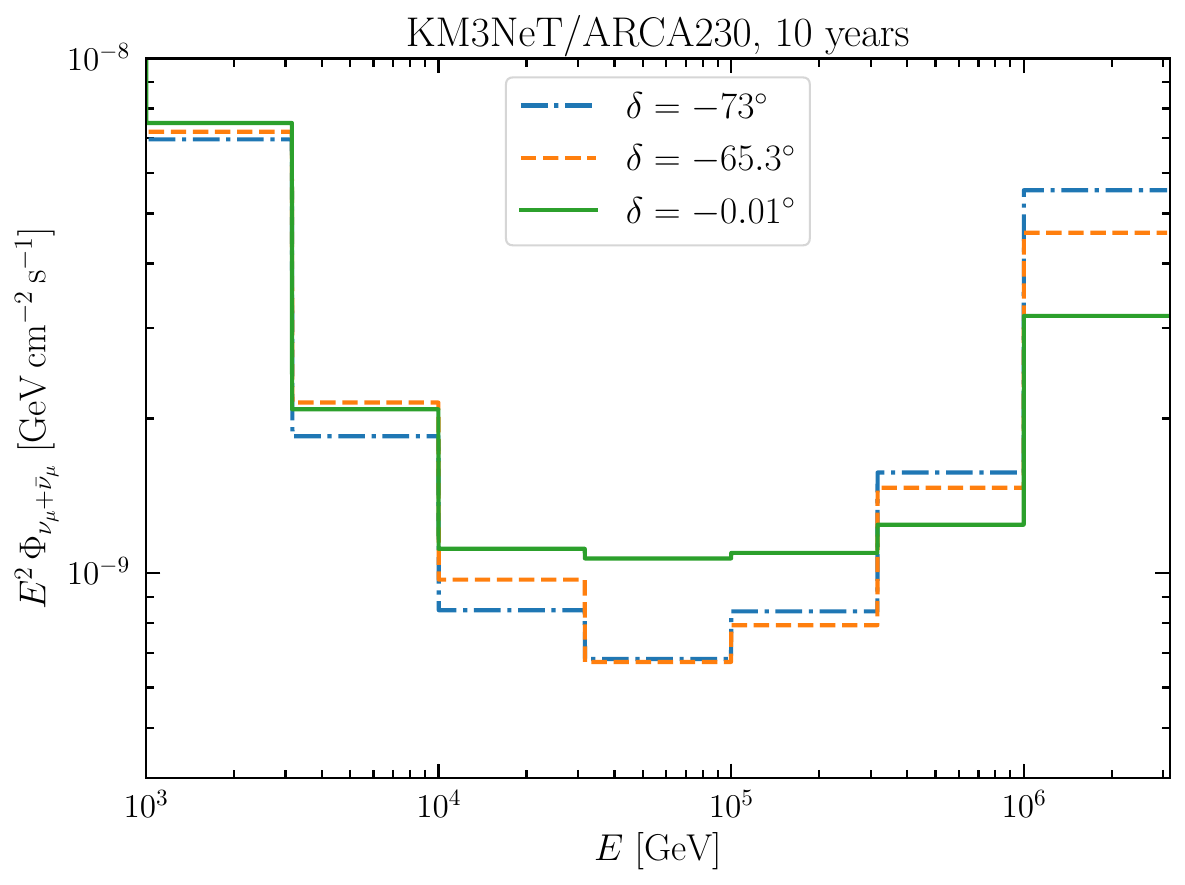}
    \caption{Comparison between point-like $90\%$ CL differential sensitivity for several declinations after 10 years of operation of the ARCA detector. The blue dashed-dotted line refers to $\delta = -73^{\circ}$, the orange dashed line to $\delta = -65^{\circ}$ and the green line to $\delta = -0.01^{\circ}$.}
    \label{fig:sensi_declination}
\end{figure}
 The sensitivities for $\delta = -73^{\circ}$ and $\delta = -65^{\circ}$ are very close as the declination bands are also near. At high energy the sensitivity deteriorates faster for $\delta = -73^{\circ}$ due to absorption from the Earth. The sensitivity at $\delta = -0.01^{\circ}$ is better at high energies due to a higher signal rate, but it is worse than the others between $10-300\, \rm TeV$ due to a higher background  rate. In fact, $\delta = -0.01^{\circ}$, the sources are only partially below the horizon (see for instance \cite{KM3NeT:2018wnd} for the ARCA detector visibility as a function of declination).

\end{document}